\documentclass[letterpaper]{article}
\usepackage{spconf,amsmath,amssymb,graphicx,booktabs,flushend,hyperref}
\usepackage{newtxtext,newtxmath}

\title{LAST: Looped Audio Spectrogram Transformer}

\name{Haider Al-Tahan$^{1}$ \quad Sean O'Brien$^{2}$ \quad
Anastasia Razdaibiedina$^{3}$\thanks{Anastasia Razdaibiedina participated in this work in an advisory capacity.} \quad N. Apurva Ratan Murty$^{1}$}
\address{$^{1}$Georgia Institute of Technology \quad
$^{2}$University of California San Diego \\
$^{3}$Google DeepMind}

\hypersetup{hidelinks,
  pdftitle={LAST: Looped Audio Spectrogram Transformer},
  pdfauthor={Haider Al-Tahan, Sean O'Brien, Anastasia Razdaibiedina, N. Apurva Ratan Murty}}
 
\begin{document}
\maketitle

\begin{abstract}
Increasing depth of transformer models improves recognition, but it comes at a substantial cost. Each additional layer requires more parameters which makes the process computationally inefficient. We ask whether additional processing can focus on integrating features already computed. Looped Audio Spectrogram Transformer (LAST) first processes all tokens, then reuses the same blocks to refine only the class token over fixed audio features, thereby making later passes inexpensive. On AudioSet, ten-pass LAST achieves 0.345 mean average precision, exceeding a twelve-layer sequential transformer by 2.1\% relative with 49.4\% fewer parameters, 42\% fewer multiply-accumulate operations, and 9.8\% higher measured throughput. Across separately trained models, increasing the pass count from two to ten improves accuracy while adding only 1.2\% computation. Further evaluations show improved robustness to temporal masking and various other auditory augmentations, with better generalization on classification tasks with music, environmental, and events sounds.
\end{abstract}


\suppressfloats[t]

\begin{figure}[t]
  \centering
  \includegraphics[width=\linewidth]{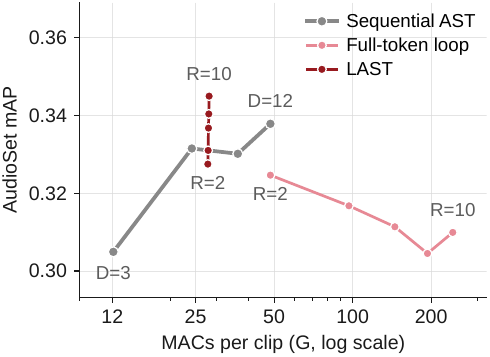}
  \caption{\textbf{LAST outperformed a deeper transformer at lower cost:} AudioSet Computation per clip (x-axis) v/s AudioSet mAP (y-axis) Sequential AST uses $D \in \{3,6,9,12\}$ layers; the recurrent models use $D=6$ shared blocks and $R \in \{2,4,6,8,10\}$ passes. }
  \label{fig:compute}
\end{figure}

\begin{figure*}[t]
  \centering
  \includegraphics[width=\textwidth]{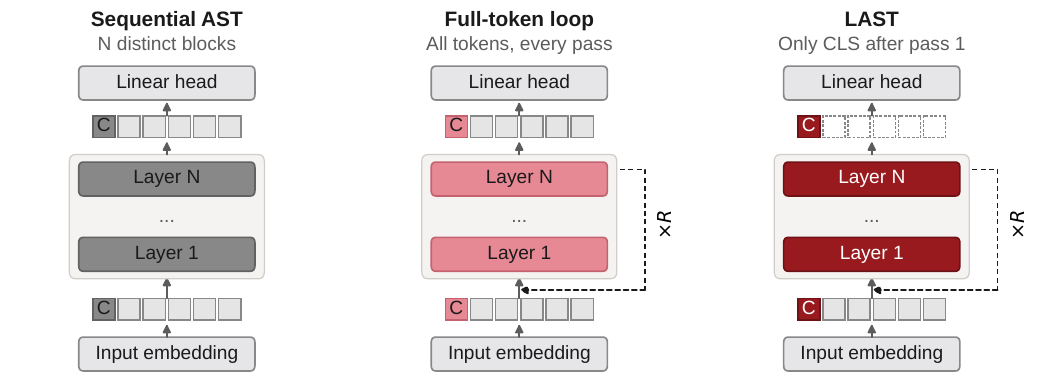}
  \caption{Sequential AST uses distinct blocks once. Full-token recurrence
  reuses the same blocks and updates all tokens on every pass. LAST uses one
  full pass, then updates only the class token using fixed audio features.
  Recurrent models use six shared blocks.}
  \label{fig:arch}
\end{figure*}

\section{Introduction}
\label{sec:intro}

Sound is a dynamic signal and recognizing audio requires integrating information across time.  Audio Spectrogram Transformers (AST) \cite{gong2021ast} integrate temporal information over layers by transforming (hundreds of) spectrogram tokens into a class-token representation. Sure enough, increasing model depth improves recognition performance to some degree but the improvement comes at a substantial computational cost. Adding layers means more parameters which makes the process inefficient. 

We wondered if there was an alterative way to improve auditory temporal integration. We know that the auditory system in human brains solve the problem of temporal integration via recurrent processing which effectively reuses parameters (neurons). And so we asked if we can we obtain the same depth \textit{effective} benefit of additional layers (depth) without increasing the number of parameters. One possibility is to use looped transformers \cite{dehghani2019universal}. Rather than progressing through distinct layers, looped transformers repeatedly apply the same transformer blocks, increasing effective depth (via recurrence) while keeping the number of parameters fixed. Recurrence has shown some benefits for tasks like reasoning in language models \cite{geiping2025recurrent,saunshi2025latent}. But sound is a more basic challenge. The signal itself unfolds over time and how recurrence contributes to audio recognition is simply not known.

LAST relates to three lines of prior work: (1) recurrence in perceptual models (2) class-token updates over fixed features, and (3) token reduction in audio transformers. Perceiver \cite{jaegle2021perceiver} iteratively cross-attends a latent array to raw image and audio inputs, and LARM \cite{kaloga2026larm} loops a speech-recognition encoder that updates every token on each pass. LAST is fundamentally different. It targets the clip-level classification and repeatedly updates \textit{only} the class token while keeping the audio features fixed after a full attention pass. Note that CaiT \cite{touvron2021layerscale} also refines a class token over fixed features but it uses \textit{separate} attention layers which require new parameters. LAST is more efficient in that it uses the model's existing encoder blocks and therefore does not add new parameters. 

We seek to adapt this method to the audio domain. Naively, each recurrent pass still updates each spectrogram token and parameter, with computational cost compared to baseline growing linearly in the number of recurrent passes. We observe that, under this naive application, the representations in for the class token tend to be less stable across recurrent steps than those of the other spectrogram tokens. Accordingly, we hypothesize that once informative audio features have been computed, additional processing may be more effectively devoted to integrating information from those pre-computed features rather than repeatedly recomputing the features themselves.

To test this hypothesis, we introduce the Looped Audio Spectrogram Transformer (LAST).
LAST first processes the full spectrogram through a set of transformer blocks to construct an audio representation. On subsequent passes, the audio tokens remain fixed while the same transformer blocks repeatedly update only the class token by attending to the fixed audio features. The audio keys and values can therefore be cached and reused across passes, making each additional loop computationally inexpensive. LAST thus increases effective depth while adding neither new parameters nor the cost of repeatedly recomputing the full audio representation.

We found that ten-pass LAST exceeds a twelve-layer sequential baseline by 2.1\% relative mAP, with 49.4\% fewer parameters and 42\% fewer MACs. Increasing the pass count improves LAST at each tested step, whereas additional full-token passes do not improve accuracy. Furthermore, we found LAST models to be more robust and generalizes better on various tasks, retaining higher accuracy on transformations such as temporal masking , and injected noise; and achieves better frozen-feature classification on benchmarks such as ESC-50 \cite{piczak2015esc50} and VGGSound \cite{chen2020vggsound} than the sequential baseline.

\section{Method}
\label{sec:method}

\subsection{Recursion through shared blocks}
\label{ssec:recurrence}

The encoder takes $N$ log-Mel spectrogram tokens and a learned class (CLS) token of width $C$, and a linear head predicts clip labels from the final class token. A sequential encoder applies $D$ distinct blocks once. A looped encoder instead reuses the same $D$ blocks for $R$ passes, yielding $DR$ block applications from $D$ sets of weights. Parameters are fixed by $D$ alone, but under full-token recurrence every token is updated on every pass and computation still scales with $R$.

\subsection{Class-token updates in LAST}
\label{ssec:cls}

LAST updates all tokens on the first pass, then fixes the audio tokens. Later passes apply attention and the feed-forward layer (MLP) only to the class token. Each block computes audio keys and values during the second pass and caches them for reuse. Attention and MLP weights are shared. Training remains end to end: gradients reach the first pass through the fixed audio features.

The cost asymmetry follows directly. A full pass costs $\mathcal{O}(D(NC^{2}+N^{2}C))$, whereas a cached class-token pass costs $\mathcal{O}(D(C^{2}+NC))$: the quadratic term in $N$ vanishes and the projection term reduces to a single query. Effective depth therefore grows with $R$ at near-constant cost. Figure~\ref{fig:arch} contrasts the three architectures, and Figure~\ref{fig:compute} the resulting accuracy--computation trade-off.

\section{Experimental setup}
\label{ssec:setup}

\textbf{Training.}  We train on the 16\,kHz archived release of AudioSet \cite{gemmeke2017audioset}, with 1{,}908{,}644 training clips. All models are trained from scratch for twelve epochs with one random seed. Inputs have 128 Mel bins and 1024 time frames, divided into $128\times2$ patches. Models use width $C=768$, twelve attention heads, an MLP expansion ratio of four, and one class token. We use AdamW with weight decay 0.01, batch size 1024, and learning rate $7.54\times10^{-5}$. The schedule has 1863 warmup steps and halves the learning rate every three epochs from epoch five. Waveform mixing is applied with probability 0.5, choosing equally between mixup \cite{zhang2018mixup} ($\alpha=10$) and temporal CutMix \cite{yun2019cutmix} ($\alpha=1$). We test sequential depths $D\in\{3,6,9,12\}$ and recurrent models with $D=6$ shared blocks and $R\in\{2,4,6,8,10\}$ passes.

\subsection{Evaluation}
\label{ssec:evaluation}

\textbf{AudioSet.} We report mean average precision (mAP) on 18{,}886 evaluation clips. Results use weights averaged models over epochs 5--12 \cite{izmailov2018swa}. We chose this shared window by mean evaluation mAP across runs. Averaging produces one model with no additional inference cost. Robustness and transfer use final-epoch checkpoints.

\textbf{Throughput.} We benchmark inference using bfloat16 and batch size 64.

\textbf{Transfer benchmarks.} We evaluate eight classification tasks: SpeechCommands V2 (SC-V2; keyword spotting) \cite{warden2018speechcommands}, VoxCeleb1 (speaker identification) \cite{nagrani2017voxceleb}, NSynth (instrument-family recognition) \cite{engel2017nsynth}, FMA-Small (music genre) \cite{defferrard2017fma}, ESC-50 \cite{piczak2015esc50} and UrbanSound8K (US8K) \cite{salamon2014urbansound} for environmental sounds, DCASE 2020 Task 1A (acoustic scenes) \cite{heittola2020tau}, and VGGSound (audio events) \cite{chen2020vggsound}. We report accuracy (\%) from one-vs-rest logistic regression with fixed hyperparameters on standardized, frozen class-token features. ESC-50 and UrbanSound8K use their official cross-validation folds. The other datasets use official train/test splits, except NSynth: we train its probe on the official validation split and evaluate on the test split to reduce training cost.

\textbf{Temporal masking.} We zero normalized spectrogram frames in regions selected by descending class-token attention or uniformly at random. Attention scores are computed separately for each model from the unmasked input's final block, averaged over heads.

\begin{figure}[t]
  \centering
  \includegraphics[width=\linewidth]{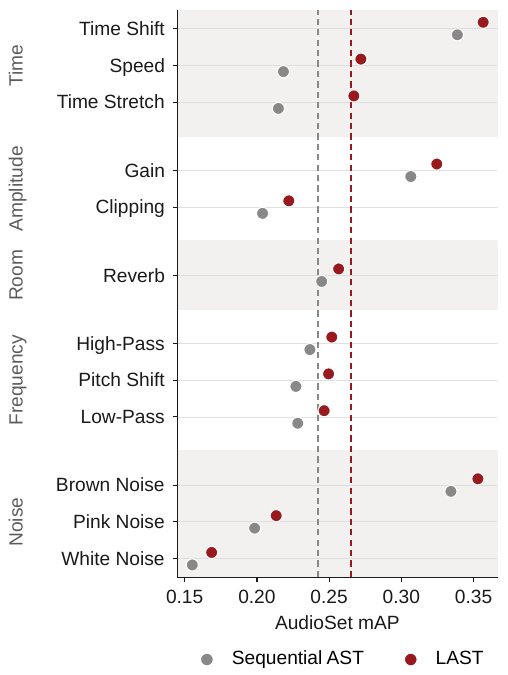}
  \caption{\textbf{LAST improved performance across the board: }AudioSet mAP under waveform corruption. Each dot is the mean over
  settings within one family; each dashed line is the mean across the twelve
  families. Models: sequential $D=12$ and LAST $D=6$, $R=10$.}
  \label{fig:signal}
\end{figure}

\textbf{Waveform corruption.} We test 38 corruption settings across twelve families (Figure~\ref{fig:signal}). These change noise, timing, frequency, reverberation, or amplitude before spectrogram extraction. Each setting uses the same 4{,}096 clips. We compare sequential $D=12$ with LAST at $D=6$, $R=10$, using final-epoch checkpoints.

\section{Results}
\label{sec:results}

\begin{figure}[t]
  \centering
  \includegraphics[width=\linewidth]{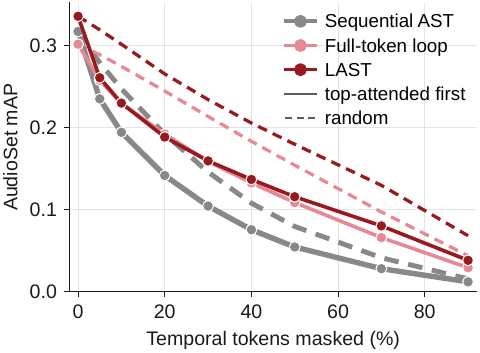}
  \caption{\textbf{LAST degraded more gracefully under temporal masking:} AudioSet mAP under temporal masking. Masking zeros the corresponding normalized spectrogram frames. Models: sequential $D=12$ and recurrent $D=6$, $R=10$.}
  \label{fig:occlusion}
\end{figure}

\subsection{Accuracy and computation}
\label{ssec:compute}

We first ask whether effective depth obtained by \textit{reusing} blocks can substitute for depth obtained by \textit{adding} them, and at what cost. Three model families are trained under an identical recipe: sequential AST at $D\in\{3,6,9,12\}$, and two recurrent variants with $D=6$ shared blocks and $R\in\{2,4,6,8,10\}$ passes. Two findings emerge. First, LAST improves at every pass count we test while its cost stays nearly flat. Second, this gain is specific to restricting later passes to the class token: full-token recurrence, given the same blocks and the same pass counts, gets \textit{worse} with depth while costing far more.

Adding blocks works, but expensively. Increasing sequential depth from $D=3$ to $D=12$ raises mAP from $0.3049$ to $0.3379$, at the cost of growing parameters from $22.3$M to $86.1$M and computation from $12.1$ to $48.4$ GMACs. Against the parameter-matched sequential $D=6$ baseline, LAST wins once $R\geq6$, and its mAP rises at every tested pass count, from $0.3275$ at $R=2$ to $0.3450$ at $R=10$. Because later passes update only the class token and reuse cached audio keys and values, these eight extra passes add just $1.2\%$ computation. The comparison against the deeper baseline is therefore favourable on all three axes at once: at $R=10$, LAST reaches $0.3450$ mAP against $0.3379$ for sequential $D=12$ --- a $2.1\%$ relative improvement --- with $49.4\%$ fewer parameters, $42\%$ fewer MACs, and $9.8\%$ higher measured throughput.

The benefit depends on how recurrence is applied. With the same six shared blocks, full-token recurrence falls from $0.3247$ mAP at $R=2$ to $0.3099$ at $R=10$, despite a fivefold increase in computation. LAST's advantage at matched $R$ grows from $0.0028$ mAP at $R=2$ to 0.0350 at $R=10$, where full-token recurrence requires approximately 8 times as many MACs. These results favor class-token recurrence over full-token recurrence under the tested training recipe. Each data point in Figure~\ref{fig:compute} is a model that is trained separately, so these results do not establish whether an already trained model benefits from additional inference passes.

\subsection{Robustness}
\label{ssec:occlusion}

Clean AudioSet mAP measures only the matched condition. If repeated class-token passes genuinely integrate evidence distributed across the clip, rather than sharpening the model's reliance on a few salient frames, the representation should degrade gracefully when that evidence is removed or distorted. We probe this from two directions: temporal masking \textit{removes} information, and waveform corruption \textit{alters} it while leaving the input complete. LAST is more robust in both, its margin over sequential AST exceeds the two models' gap on clean audio, and on corruptions it leads in every family tested.

\textbf{Temporal masking.} We progressively zero normalized spectrogram frames, selecting regions either by descending class-token attention or uniformly at random. LAST retains more accuracy than sequential AST under both attention-based and random temporal masking (Figure~\ref{fig:occlusion}). It achieves the highest mean mAP among the three models. Integrating each curve over 0--90\% masking and dividing by that range gives mean mAP of 0.1383 versus 0.0905 for LAST and sequential AST under attention-based masking, and 0.1970 versus 0.1187 under random masking.

\textbf{Waveform corruption.}Masking withholds information outright; waveform corruption instead perturbs the signal before spectrogram extraction, so the model receives a complete but distorted input. Across $38$ settings spanning twelve families --- noise, timing, frequency, reverberation, and amplitude --- LAST attains higher mAP in every family (Figure~\ref{fig:signal}). Averaging within each family and then across families gives $0.2651$ for LAST against $0.2422$ for sequential $D=12$, a gap of $0.0229$. The largest gaps occur under speed perturbation and time stretching, compared with 0.012 to 0.023 across the remaining families.

\subsection{Frozen-feature transfer}
\label{ssec:transfer}

\begin{table}[t]
\centering
\caption{\textbf{LAST's features transferred better on most tasks, with half the parameters:} Frozen-feature classification accuracy (\%) using final-epoch encoders and the protocols in Section~\ref{ssec:evaluation}.}
\label{tab:downstream}
\begin{tabular}{l r r r}
\toprule
 & Sequential & Looped, & \textbf{LAST} \\
 & AST & full-token &  \\
\midrule
Depth & 12 & 6 & 6 \\
Recurrences & -- & 2 & 10 \\
Parameters (M) & 86.1 & 43.5 & 43.5 \\
MACs (G) & 48.4 & 48.4 & \textbf{28.2} \\
Throughput (clips/s) & 571 & 572 & \textbf{628} \\
\midrule
AudioSet mAP & 0.3379 & 0.3247 & \textbf{0.3450} \\
\midrule
\multicolumn{4}{l}{\textit{Voice \& Speech}} \\
\quad SC-V2 & \textbf{71.9} & 71.6 & 69.8 \\
\quad VoxCeleb & 3.6 & \textbf{4.3} & 4.0 \\
\midrule
\multicolumn{4}{l}{\textit{Music}} \\
\quad NSynth & \textbf{96.9} & 96.1 & 96.8 \\
\quad  FMA & 43.5 & 42.1 & \textbf{50.9} \\
\midrule
\multicolumn{4}{l}{\textit{Environmental \& Scenes}} \\
\quad ESC-50 & 84.1 & 83.6 & \textbf{85.7} \\
\quad US8K & \textbf{78.1} & 76.8 & 77.8 \\
\quad DCASE & 46.7 & \textbf{48.6} & 48.3 \\
\quad VGGSound & 34.9 & 34.6 & \textbf{36.3} \\
\bottomrule
\end{tabular}
\end{table}

Transfer-set benchmarks asks a different question: is the class token, having been shaped by repeated integration, a better general-purpose clip descriptor? We freeze the class-token features and fit one-vs-rest logistic regression with fixed hyperparameters on eight tasks spanning speech, music, environmental sound, and acoustic scenes (Table~\ref{tab:downstream}). LAST outperforms sequential $D=12$ on five of eight transfer benchmarks. The largest gain is on FMA-Small, from 43.5\% to 50.9\%; LAST also improves ESC-50 and VGGSound accuracy. DCASE accuracy rises from 46.7\% to 48.3\%.

\clearpage
\bibliographystyle{IEEEbib}
\bibliography{refs}

\begin{thebibliography}{10}

\bibitem{gong2021ast}
Yuan Gong, Yu-An Chung, and James Glass,
\newblock ``{AST}: Audio spectrogram transformer,''
\newblock in {\em Proc. Interspeech}, 2021, pp. 571--575.

\bibitem{dehghani2019universal}
Mostafa Dehghani, Stephan Gouws, Oriol Vinyals, Jakob Uszkoreit, and Lukasz
  Kaiser,
\newblock ``Universal transformers,''
\newblock in {\em Proc. Int. Conf. Learning Representations (ICLR)}, 2019.

\bibitem{geiping2025recurrent}
Jonas Geiping, Sean McLeish, Neel Jain, John Kirchenbauer, Siddharth Singh,
  Brian Bartoldson, Bhavya Kailkhura, Abhinav Bhatele, and Tom Goldstein,
\newblock ``Scaling up test-time compute with latent reasoning: A recurrent
  depth approach,''
\newblock in {\em Advances in Neural Information Processing Systems (NeurIPS)},
  2025, vol.~38, pp. 41340--41391.

\bibitem{saunshi2025latent}
Nikunj Saunshi, Nishanth Dikkala, Zhiyuan Li, Sanjiv Kumar, and Sashank~J.
  Reddi,
\newblock ``Reasoning with latent thoughts: On the power of looped
  transformers,''
\newblock in {\em Proc. Int. Conf. Learning Representations (ICLR)}, 2025.

\bibitem{jaegle2021perceiver}
Andrew Jaegle, Felix Gimeno, Andy Brock, Andrew Zisserman, Oriol Vinyals, and
  Joao Carreira,
\newblock ``Perceiver: General perception with iterative attention,''
\newblock in {\em Proc. Int. Conf. Machine Learning (ICML)}, 2021.

\bibitem{kaloga2026larm}
Yacouba Kaloga, Shashi Kumar, Shakeel~A. Sheikh, Driss Khalil, Petr Motlicek,
  and Ina Kodrasi,
\newblock ``Test-time compute scaling for {ASR} with depth-conditioned looped
  transformers,''
\newblock {\em arXiv preprint arXiv:2606.04678}, 2026.

\bibitem{touvron2021layerscale}
Hugo Touvron, Matthieu Cord, Alexandre Sablayrolles, Gabriel Synnaeve, and
  Herv{\'e} J{\'e}gou,
\newblock ``Going deeper with image transformers,''
\newblock in {\em Proc. IEEE/CVF Int. Conf. Computer Vision (ICCV)}, 2021.

\bibitem{piczak2015esc50}
Karol~J. Piczak,
\newblock ``{ESC}: Dataset for environmental sound classification,''
\newblock in {\em Proc. ACM Int. Conf. Multimedia}, 2015, pp. 1015--1018.

\bibitem{chen2020vggsound}
Honglie Chen, Weidi Xie, Andrea Vedaldi, and Andrew Zisserman,
\newblock ``{VGGSound}: A large-scale audio-visual dataset,''
\newblock in {\em Proc. IEEE ICASSP}, 2020, pp. 721--725.

\bibitem{gemmeke2017audioset}
Jort~F. Gemmeke, Daniel P.~W. Ellis, Dylan Freedman, Aren Jansen, Wade
  Lawrence, R.~Channing Moore, Manoj Plakal, and Marvin Ritter,
\newblock ``Audio set: An ontology and human-labeled dataset for audio
  events,''
\newblock in {\em Proc. IEEE ICASSP}, 2017, pp. 776--780.

\bibitem{zhang2018mixup}
Hongyi Zhang, Moustapha Cisse, Yann~N. Dauphin, and David Lopez-Paz,
\newblock ``mixup: Beyond empirical risk minimization,''
\newblock in {\em Proc. Int. Conf. Learning Representations (ICLR)}, 2018.

\bibitem{yun2019cutmix}
Sangdoo Yun, Dongyoon Han, Seong~Joon Oh, Sanghyuk Chun, Junsuk Choe, and
  Youngjoon Yoo,
\newblock ``{CutMix}: Regularization strategy to train strong classifiers with
  localizable features,''
\newblock in {\em Proc. IEEE/CVF Int. Conf. Computer Vision (ICCV)}, 2019.

\bibitem{izmailov2018swa}
Pavel Izmailov, Dmitrii Podoprikhin, Timur Garipov, Dmitry Vetrov, and
  Andrew~Gordon Wilson,
\newblock ``Averaging weights leads to wider optima and better
  generalization,''
\newblock in {\em Proc. Conf. Uncertainty in Artificial Intelligence (UAI)},
  2018.

\bibitem{warden2018speechcommands}
Pete Warden,
\newblock ``Speech commands: A dataset for limited-vocabulary speech
  recognition,''
\newblock {\em arXiv preprint arXiv:1804.03209}, 2018.

\bibitem{nagrani2017voxceleb}
Arsha Nagrani, Joon~Son Chung, and Andrew Zisserman,
\newblock ``{VoxCeleb}: A large-scale speaker identification dataset,''
\newblock in {\em Proc. Interspeech}, 2017, pp. 2616--2620.

\bibitem{engel2017nsynth}
Jesse Engel, Cinjon Resnick, Adam Roberts, Sander Dieleman, Mohammad Norouzi,
  Douglas Eck, and Karen Simonyan,
\newblock ``Neural audio synthesis of musical notes with {WaveNet}
  autoencoders,''
\newblock in {\em Proc. Int. Conf. Machine Learning (ICML)}, 2017.

\bibitem{defferrard2017fma}
Micha{\"e}l Defferrard, Kirell Benzi, Pierre Vandergheynst, and Xavier Bresson,
\newblock ``{FMA}: A dataset for music analysis,''
\newblock in {\em Proc. Int. Society for Music Information Retrieval Conf.
  (ISMIR)}, 2017.

\bibitem{salamon2014urbansound}
Justin Salamon, Christopher Jacoby, and Juan~Pablo Bello,
\newblock ``A dataset and taxonomy for urban sound research,''
\newblock in {\em Proc. ACM Int. Conf. Multimedia}, 2014, pp. 1041--1044.

\bibitem{heittola2020tau}
Toni Heittola, Annamaria Mesaros, and Tuomas Virtanen,
\newblock ``Acoustic scene classification in {DCASE} 2020 challenge:
  Generalization across devices and low complexity solutions,''
\newblock in {\em Proc. Detection and Classification of Acoustic Scenes and
  Events Workshop (DCASE)}, 2020, pp. 56--60.

\end{thebibliography}

\end{document}